\documentstyle[sprocl]{article}
\def\Journal#1#2#3#4{{#1} {\bf #2}, #3 (#4)}
\begin{document}
\title{DEVELOPMENT OF THE SOLAR TOWER ATMOSPHERIC
CHERENKOV EFFECT EXPERIMENT (STACEE)}
\author{ Ren\'e A. Ong }
\address{The Enrico Fermi Institute,
The University of Chicago, \\
5640 S. Ellis Ave, Chicago, IL 60637, USA}
\author{(For the STACEE Collaboration \cite{Collaboration})}
\address{}
\maketitle\abstracts{
STACEE is a proposed telescope for ground-based gamma-ray
astrophysics between 25 and 500 GeV.
The telescope will make use of large mirrors available
at a solar research facility to achieve an energy threshold lower
than existing ground-based instruments.
This paper describes recent development work on STACEE.}

\section{Introduction}

Discoveries from the Compton Gamma Ray Observatory (CGRO) \cite{EGRET} 
and from ground-based experiments \cite{Whipple}
indicate that the high energy sky is rich with interesting astrophysics.
Yet, there is a gap in experimental coverage between 20 and 250 GeV.
Satellite instruments, such as GLAST,\cite{GLAST} may eventually
extend their reach above 20 GeV, but the experiments with the
most promise to explore the gap in the near future are ground-based
detectors using the atmospheric Cherenkov technique.

The energy threshold of atmospheric Cherenkov detectors
is governed by a number of parameters,
of which the easiest to control is the mirror collection
area.\cite{Weekes}
Large collection area translates into lower energy threshold,
and large solar mirrors (heliostats)
are readily available at existing power facilities.
Since early 1994, we have been developing an
experiment (STACEE) to use heliostat mirrors for Cherenkov astronomy.
A similar experiment (CELESTE)
is also under development in France.\cite{Celeste}

\section{STACEE Development}

Our development work has concentrated on the key issues
associated with building an innovative atmospheric Cherenkov detector.
We have carried out tests using prototype detector equipment
at two solar heliostat fields,
the Solar Two Power Plant (Barstow, CA) and the National Solar
Thermal Test Facility (NSTTF) at Sandia National Laboratories 
(Albuquerque, NM).
The results from work at Solar Two have been published \cite{Ong} and
in 1996, the successful tests at the NSTTF encouraged us to develop
a complete instrument design using 48 heliostats at Sandia.
{\bf Documents describing the Test Results and the STACEE design
can be found on the Web.}\cite{WWW}\ \
Here we {\em very briefly} summarize these documents.

\section{Results from the Sandia Tests}

We carried out two tests at Sandia (Aug. and Oct. 1996).
To summarize:
\begin{itemize}
\item 
we verified that the site is suitable
for Cherenkov astronomy by measuring the clarity of the sky
and the ambient flux of night sky photons, and
\item we determined that the heliostats
have excellent pointing accuracy ($\sim 0.04^\circ$) and
stability ($\sim 0.05^\circ$), and typical
spot sizes of $1.5\,$m
and reflectivities of $\sim 80$\%.
\end{itemize}

\noindent We built a complete detector prototype consisting of
a $2\,$m secondary mirror, support structure, photomultiplier tube (PMT)
camera, electronics and data acquisition system.\cite{WWW}
The detector prototype performed extremely well, and it proved easy
to detect atmospheric Cherenkov radiation 
from cosmic ray showers with little accidental background.
Using these showers:
\begin{itemize}
\item we measured the trigger rate
dependence on zenith angle, the effect of tilting the
heliostats to the interaction point, and the cosmic
cosmic ray spectral index, and
\item
from the trigger rate (5 Hz) and simulations, we determined 
a cosmic ray energy threshold of $\sim 290\,$GeV for vertical showers.
\end{itemize}
The cosmic ray threshold translates into an effective gamma-ray
threshold of $\sim 75\,$GeV, indicating that
the prototype instrument operated at a lower energy threshold
than any atmospheric Cherenkov detector to date.

\section{Overall Detector Design}

The full experiment will use 48 heliostats at Sandia,
corresponding to a total mirror area of $\sim 1770\,$m$^2$.
The heliostats will be divided into three sectors and 
Cherenkov light from each sector will be reflected onto a separate
$2\,$m diameter secondary mirror.
Each secondary will image the light onto a
16-element camera, consisting of PMTs equipped with Winston
cones.

The PMT signals will be amplified and discriminated.
The discriminated signals will be delayed and combined
to form an overall multiplicity trigger.  
The amplified PMT signals will be continuously sampled by a
digitizer which will store a waveform for each PMT upon
receipt of a trigger.
The PMT arrival times and pulse-heights 
will be determined from the digitized waveforms.

Simulations show that STACEE will have a substantial collection area
($10,500\,$m$^2$) for 50 GeV gamma-ray primaries, and
that the experiment will be fully efficient by $75\,$GeV.
In addition, the experiment should possess substantial capability
to reject hadronic cosmic rays (rejection factor of $\sim$210 at
50 GeV and $\sim 95$ at 100 GeV), as a result of the rapid decease
in the Cherenkov yield for cosmic rays below 200 GeV, the narrow
field-of-view of each heliostat, the multiplicity trigger condition,
and the measured lateral distribution of the Cherenkov light.
From simulations we expect that STACEE will have excellent point
source sensitivity ($\sim 8\sigma$ significance on the Crab in one hour).

\section{Summary}

We have completed the design of an innovative atmospheric Cherenkov
detector sensitive to gamma-rays in an unexplored energy region.
The complete experiment can be built on a two
year timescale.

\section*{Acknowledgments} 
We thank Richard Fernholz,
G.H. Marion, Antonino Miceli,
Heather Ueunten,
Patrick Fleury, Eric Par\'e, David Smith, and
staff of the National Solar Thermal Test
Facility.
This work was supported by the National Science Foundation,
the Natural Sciences and Engineering Research Council, the
California Space Institute, and the University of Chicago.

\end{document}